\documentclass[aps,prc,twocolumn,showpacs,superscriptaddress,showpacs]{revtex4-1}
\usepackage{CJK}
\usepackage{graphicx}
\usepackage{dcolumn}
\usepackage{bm}
\usepackage{float}
\usepackage{booktabs}
\usepackage{amsmath}
\usepackage{color}
\usepackage{array, longtable}
\usepackage{multirow}
\usepackage{hhline}
\usepackage{color,hyperref}
\usepackage{threeparttable} 
\usepackage{mathtools}
\hypersetup{colorlinks,breaklinks,linkcolor=blue,urlcolor=blue,anchorcolor=blue,citecolor=blue}
\begin{document}
\title{Supplementary Material for \\ "Evidence against  the wobbling nature of low-spin bands in $^{135}$Pr" }

\author{B. F. Lv}
\affiliation{Key Laboratory of High Precision Nuclear Spectroscopy and Center for Nuclear Matter Science, Institute of Modern Physics, Chinese Academy of Sciences, Lanzhou 730000, People's Republic of China} 
\affiliation{School of Nuclear Science and Technology, University of Chinese Academy of Science, Beijing 100049, People's Republic of China}
 \author{C. M. Petrache}
 \email[Corresponding author: ]{ petrache@ijclab.in2p3.fr}
\affiliation{Universit\'{e} Paris-Saclay, CNRS/IN2P3, IJCLab, 91405 Orsay, France}
\author{E. A. Lawrie}
\affiliation{iThemba LABS, National Research Foundation, PO Box 722, Somerset West 7129, South Africa}
\affiliation{Department of Physics, University of the Western Cape, Private Bag X17, 7535 Bellville, South Africa}
 \author{S. Guo}
\affiliation{Key Laboratory of High Precision Nuclear Spectroscopy and Center for Nuclear Matter Science, Institute of Modern Physics, Chinese Academy of Sciences, Lanzhou 730000, People's Republic of China} 
\affiliation{School of Nuclear Science and Technology, University of Chinese Academy of Science, Beijing 100049, People's Republic of China}
\author{A. Astier}
\affiliation{Universit\'{e} Paris-Saclay, CNRS/IN2P3, IJCLab, 91405 Orsay, France}
\author{E. Dupont}
\affiliation{Universit\'{e} Paris-Saclay, CNRS/IN2P3, IJCLab, 91405 Orsay, France}
 \author{K. K. Zheng} 
 \affiliation{Universit\'{e} Paris-Saclay, CNRS/IN2P3, IJCLab, 91405 Orsay, France}
\affiliation{Key Laboratory of High Precision Nuclear Spectroscopy and Center for Nuclear Matter Science, Institute of Modern Physics, Chinese Academy of Sciences, Lanzhou 730000, People's Republic of China} 
\affiliation{School of Nuclear Science and Technology, University of Chinese Academy of Science, Beijing 100049, People's Republic of China}
\author{H. J. Ong}
\affiliation{Key Laboratory of High Precision Nuclear Spectroscopy and Center for Nuclear Matter Science, Institute of Modern Physics, Chinese Academy of Sciences, Lanzhou 730000, People's Republic of China} 
\affiliation{School of Nuclear Science and Technology, University of Chinese Academy of Science, Beijing 100049, People's Republic of China}
\author{J. G. Wang}
\affiliation{Key Laboratory of High Precision Nuclear Spectroscopy and Center for Nuclear Matter Science, Institute of Modern Physics, Chinese Academy of Sciences, Lanzhou 730000, People's Republic of China} 
\affiliation{School of Nuclear Science and Technology, University of Chinese Academy of Science, Beijing 100049, People's Republic of China}
\author{X. H. Zhou}
\affiliation{Key Laboratory of High Precision Nuclear Spectroscopy and Center for Nuclear Matter Science, Institute of Modern Physics, Chinese Academy of Sciences, Lanzhou 730000, People's Republic of China} 
\affiliation{School of Nuclear Science and Technology, University of Chinese Academy of Science, Beijing 100049, People's Republic of China}
\author{Z. Y. Sun}
\affiliation{Key Laboratory of High Precision Nuclear Spectroscopy and Center for Nuclear Matter Science, Institute of Modern Physics, Chinese Academy of Sciences, Lanzhou 730000, People's Republic of China} 
\affiliation{School of Nuclear Science and Technology, University of Chinese Academy of Science, Beijing 100049, People's Republic of China}
 \author{P. Greenlees}
 \affiliation{Department of Physics, University of Jyv\"askyl\"a, Jyv\"askyl\"a FIN-40014, Finland}
 \author{H. Badran}
\affiliation{Department of Physics, University of Jyv\"askyl\"a, Jyv\"askyl\"a FIN-40014, Finland}
\author{T. Calverley}  
 \affiliation{Department of Physics, University of Jyv\"askyl\"a, Jyv\"askyl\"a FIN-40014, Finland}
\affiliation{Department of Physics, University of Liverpool, The Oliver Lodge Laboratory, Liverpool L69 7ZE, United Kingdom}
\author{D. M. Cox}
 \affiliation{Department of Physics, University of Jyv\"askyl\"a, Jyv\"askyl\"a FIN-40014, Finland}
  \author{T. Grahn}
 \affiliation{Department of Physics, University of Jyv\"askyl\"a, Jyv\"askyl\"a FIN-40014, Finland}
 \author{J. Hilton}  
 \affiliation{Department of Physics, University of Jyv\"askyl\"a, Jyv\"askyl\"a FIN-40014, Finland}
 \affiliation{Department of Physics, University of Liverpool, The Oliver Lodge Laboratory, Liverpool L69 7ZE, United Kingdom}
\author{R. Julin}
 \affiliation{Department of Physics, University of Jyv\"askyl\"a, Jyv\"askyl\"a FIN-40014, Finland}
 \author{S. Juutinen}
  \affiliation{Department of Physics, University of Jyv\"askyl\"a, Jyv\"askyl\"a FIN-40014, Finland}
\author{J. Konki}
 \affiliation{Department of Physics, University of Jyv\"askyl\"a, Jyv\"askyl\"a FIN-40014, Finland}
\author{J. Pakarinen}
 \affiliation{Department of Physics, University of Jyv\"askyl\"a, Jyv\"askyl\"a FIN-40014, Finland}
\author{P. Papadakis}
 \affiliation{Department of Physics, University of Jyv\"askyl\"a, Jyv\"askyl\"a FIN-40014, Finland}
 \affiliation{STFC Daresbury Laboratory, Daresbury, Warrington, WA4 4AD, UK}
\author{J. Partanen}
 \affiliation{Department of Physics, University of Jyv\"askyl\"a, Jyv\"askyl\"a FIN-40014, Finland}
\author{P. Rahkila}
 \affiliation{Department of Physics, University of Jyv\"askyl\"a, Jyv\"askyl\"a FIN-40014, Finland}
\author{P. Ruotsalainen}
 \affiliation{Department of Physics, University of Jyv\"askyl\"a, Jyv\"askyl\"a FIN-40014, Finland}
 \author{M. Sandzelius}
  \affiliation{Department of Physics, University of Jyv\"askyl\"a, Jyv\"askyl\"a FIN-40014, Finland}
\author{J. Saren}
 \affiliation{Department of Physics, University of Jyv\"askyl\"a, Jyv\"askyl\"a FIN-40014, Finland}
\author{C. Scholey}
 \affiliation{Department of Physics, University of Jyv\"askyl\"a, Jyv\"askyl\"a FIN-40014, Finland}
\author{J. Sorri}
 \affiliation{Department of Physics, University of Jyv\"askyl\"a, Jyv\"askyl\"a FIN-40014, Finland}
 \affiliation{Sodankyl\"a Geophysical Observatory, University of Oulu, FIN-99600 Sodankyl\"a, Finland} 
\author{S. Stolze}
 \affiliation{Department of Physics, University of Jyv\"askyl\"a, Jyv\"askyl\"a FIN-40014, Finland}
 \author{J. Uusitalo} 
  \affiliation{Department of Physics, University of Jyv\"askyl\"a, Jyv\"askyl\"a FIN-40014, Finland}
 \author{B. Cederwall}
 \affiliation {KTH Department of Physics,S-10691 Stockholm, Sweden}
 \author{A. Ertoprak}
 \affiliation {KTH Department of Physics,S-10691 Stockholm, Sweden}
\author{H. Liu}  
 \affiliation {KTH Department of Physics,S-10691 Stockholm, Sweden}
\affiliation{Institute of Modern Physics, Chinese Academy of Sciences, Lanzhou 730000, China}
\author{I. Kuti}
\affiliation{Institute for Nuclear Research (Atomki), Pf. 51, 4001 Debrecen, Hungary}
\author{J. Tim\'ar}
\affiliation{Institute for Nuclear Research (Atomki), Pf. 51, 4001 Debrecen, Hungary}
\author{A. Tucholski}
\affiliation{University of Warsaw, Heavy Ion Laboratory, Pasteura 5a, 02-093 Warsaw, Poland}  
\author{J. Srebrny}
\address{University of Warsaw, Heavy Ion Laboratory, Pasteura 5a, 02-093 Warsaw, Poland}  
\author{C. Andreoiu}
\affiliation{Department of Chemistry, Simon Fraser University, Burnaby, BC V5A 1S6, Canada}

\maketitle
\section{Experimental details}
In present work, spin and parity assigned to the low-lying states in $^{135}$Pr were based on the measurement of  the angular distribution ratios (anisotropy) ratios $R_{ac}$ \cite{KramerFlecken1989333,Chiara.75.054305} and linear polarization $P$~\cite{STAROSTA199916}. Particularly, the mixing ratios of the $\Delta I$=1 transitions of sufficient intensity have been extracted. This approach is similar to the polarization direction correlation method (PDCO)~\cite{PDCOStarosta1999, KramerFlecken1989333, PDCODroste1996}, but extracts the information from spectra with much higher statistics because it involves more detectors grouped around two angles. 

The values of  $R_{ac}$ were extracted from two $\gamma-\gamma$ matrices, one having $\gamma$ rays detected by the Phase 1 tapered detectors at ($157.6^{\circ}$  and  $133.6^{\circ}$) on one axis  and  by all detectors on the other axis, and a second matrix having $\gamma$ rays detected by the clover detectors around  $90^{\circ}$  ($75.5^{\circ}$ and $104.5^{\circ}$) on one axis and by all detectors on the other axis. The same energy gates on the axis of all the detectors were put in both matrices, and gated spectra were projected on the other axis.  $R_{ac} $ ratios were calculated  using the formula
 \begin{equation} 
   R_{ac} =\frac {I_{\gamma} (157.6^{\circ}+133.6^{\circ},~\rm{gated~on~all~angles})}{I_{\gamma}  (\approx 90^{\circ},~\rm{gated~on~all~angles)}}.
\end{equation}
The expected $R_{ac}$ ratios are $\approx$ 0.8  and $\approx$ 1.4 for stretched dipole and quadrupole transitions, respectively. 

The linear polarization analysis was performed only for the 747-, 813- and 450-keV transitions, as described in Refs. \cite{PhysRevC.92.044310}. Two matrices were constructed with events in which $\gamma$-rays were scattered between the crystals of a clover detector in parallel (perpendicular) directions relative to the beam on one axis, and $\gamma$-rays detected by all detectors on the other axis. The linear polarization values were then derived according to 
 \begin{equation} 
  P ={1 \over Q(E_{\gamma})} \frac {a(E_{\gamma})N_\perp- N_\parallel}{a(E_{\gamma})N_\perp+ N_\parallel},
\end{equation}
where $N_\perp$ and $N_\parallel$ denote the number of coincidence events for a $\gamma$-ray of interest obtained by setting the same gates in the two  asymmetric matrices on the all-detector axis.  The $a(E_\gamma$)  denotes the normalization factor due to the asymmetry in the response of the perpendicular and parallel clover segments. In the present work,  $a(E_\gamma)$ = 0.9844(21) + 2(1)$\times10^{-5} E_\gamma$, where $E_\gamma$ is the transition energy. The $Q(E_{\gamma})$ dependence of the polarization sensitivity of the clover detectors working as polarimeters in the JUROGAM II array was adopted from the measurements reported in Ref.~\cite{PhysRevC.92.044310}.  Detailed experimental information on levels and  $\gamma$-ray transitions is given in Table \ref{tab2}. 

\begin{figure*}[!htbp]
 \centering
\vskip -.cm
\rotatebox{-90}{\scalebox{0.62}{\includegraphics[width=\textwidth]{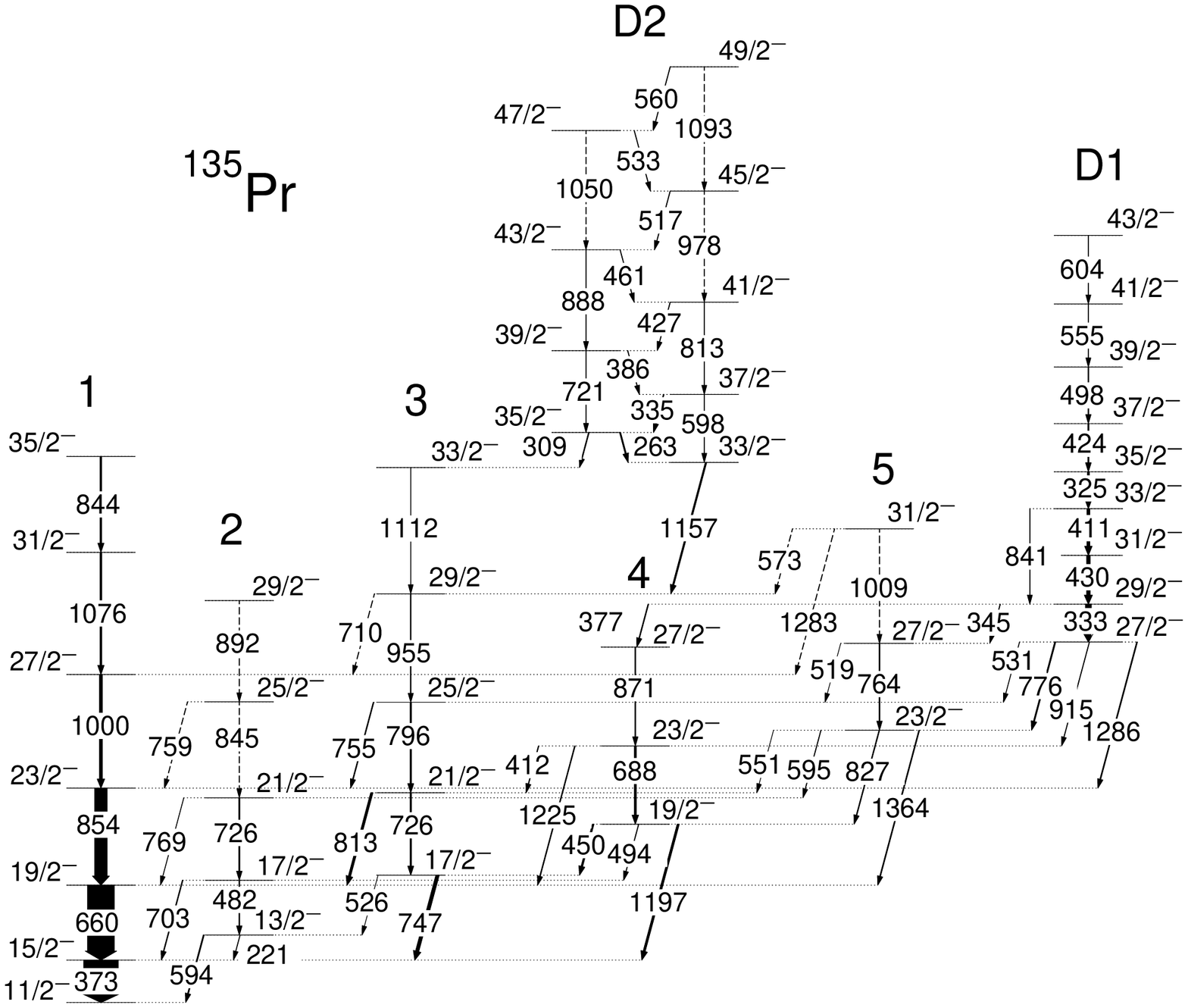}}}
\vskip .cm
\caption []{\label{fig1s}The negative-parity level scheme of $^{135}$Pr obtained from present work are shown with solid lines, while those reported in previous works~\cite{135pr-2015PRL, 135pr-2019PLB} but not observed in the present data are drawn with dashed lines.}
\end{figure*}
\begin{figure}[]
 \centering
\vskip -.cm
\rotatebox{-0}{\scalebox{0.65}{\includegraphics[width=\textwidth]{fig2s.eps}}}
\vskip .cm
\caption []{\label{fig2s}(Color online) Representative double-gated spectra  showing the low-lying structure of $^{135}$Pr. (a) Double gated on the 747- and 450-keV transitions. (b) Double gated on the 450- and 688-keV transitions. (c) Double gated on the 813- and 796-keV transition.}
\end{figure}
\begin{figure}[!htbp]
 \centering
\vskip -.cm
\rotatebox{-0}{\scalebox{0.48}{\includegraphics[width=\textwidth]{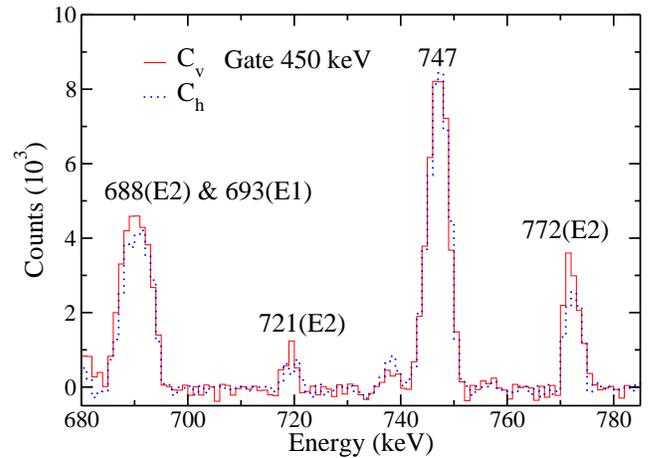}}}
\vskip .cm
\caption []{\label{fig3s}(Color online) Polarization spectra for the 747-keV transition of $^{135}$Pr in which the perpendicular (C$_v$) and parallel (C$_h$) spectra are marked with red line and blue dashed-line, respectively.}
\end{figure}
\begin{figure}[!htbp]
 \centering
\vskip -.cm
\rotatebox{-0}{\scalebox{0.48}{\includegraphics[width=\textwidth]{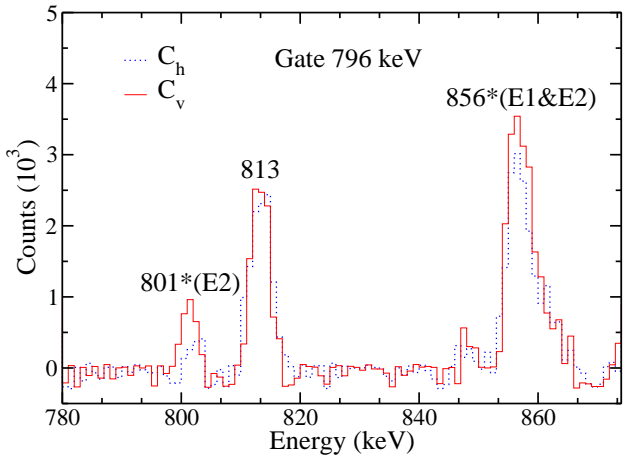}}}
\vskip .cm
\caption []{\label{fig4s}(Color online) Polarization spectra for the 813-keV transition in which the perpendicular (C$_V$ ) and parallel (C$_h$) spectra are marked with red and blue-dashed lines, respectively. The 801-keV ($E2$) and 856-keV ($E1$) transitions belong to $^{137}$Nd. The  856-keV peak has also a component from HD1 band in $^{136}$Nd.}
\end{figure}
\begin{figure}[!htbp]
 \centering
\vskip -.cm
\rotatebox{-0}{\scalebox{0.48}{\includegraphics[width=\textwidth]{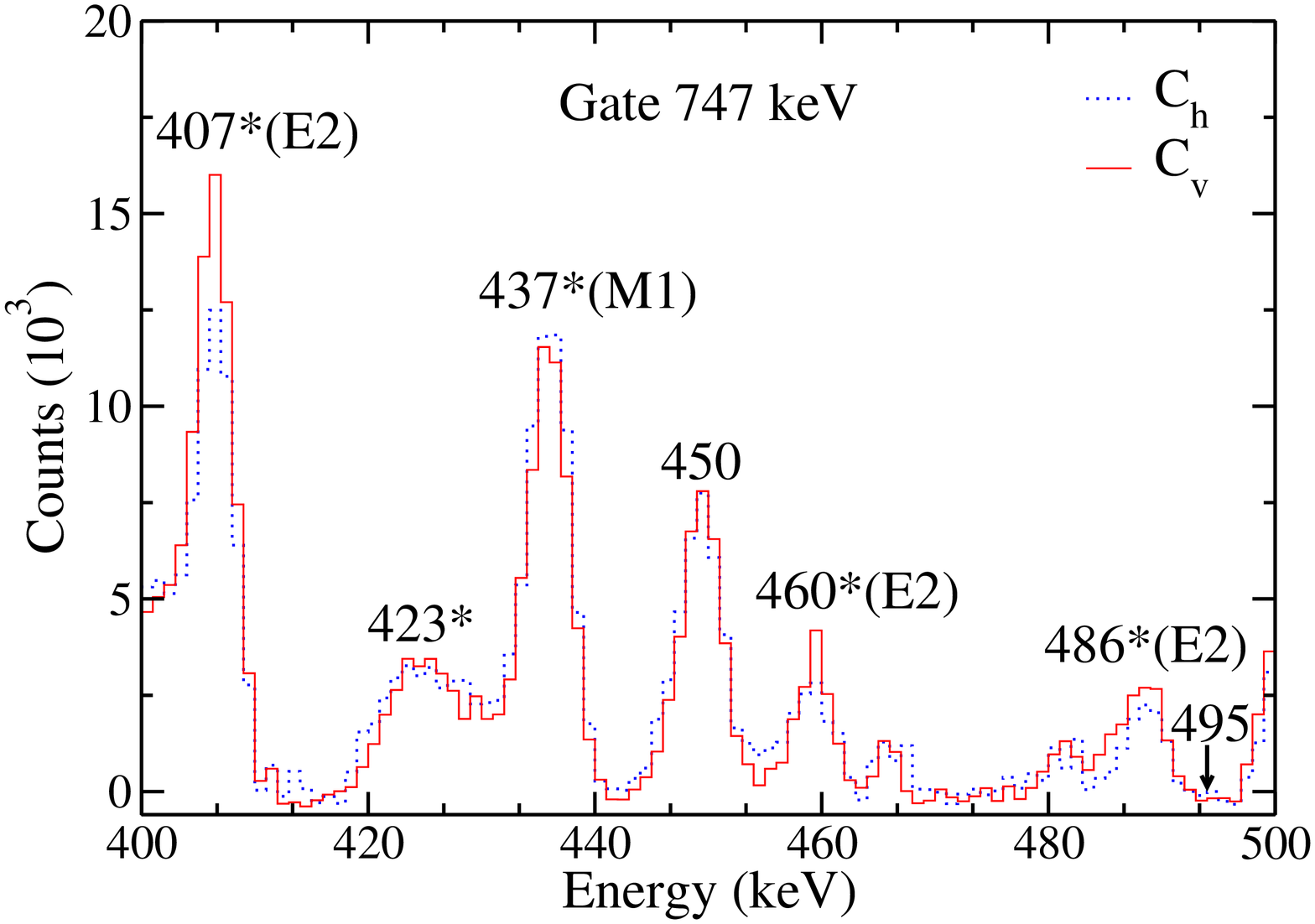}}}
\vskip .cm
\caption []{\label{fig5s}(Color online) Polarization spectra measured for the 450-keV transition in which the perpendicular (C$_V$ ) and parallel (C$_h$) spectra are marked with red and blue-dashed  lines, respectively.  The 407-keV ($E2$), 737-keV ($M1$), and 460-keV ($E2$) transitions are present in the spectrum due to the strong 749-keV contaminant transition from $^{137}$Nd. More details are given in the text.}
\end{figure}
In order to obtain reliable results for the linear polarization asymmetries, in the first step, it is of critical importance to ensure that the used gates are clean and the transition to be analyzed is strong.  Note that the polarization asymmetries are extracted from single-gated spectra obtained from $\gamma\gamma$-coincidence matrices, in which transitions from other nuclei can be present. It is therefore extremely important to avoid the gates which show coincidences with transitions having energies similar to the transitions of interest. We have therefore carefully checked the coincidence relationships and, after identifying the possible contaminations, we chose the cleanest spectra.

The partial level scheme of $^{135}$Pr including the bands discussed in the present manuscript, as well as the two dipole bands D1 and D2 which decay to them, which are needed to explain the presence of certain peaks in Fig. \ref{fig2s}, is given in Fig. \ref{fig1s}.

Representative double-gated $\gamma$-ray spectra showing the low-lying negative-parity bands are presented in Fig.~\ref{fig2s}. In panel (a) of Fig.~\ref{fig2s}, in which the spectrum obtained by double-gating on the 450- and 747-keV transitions is shown, one can observe the strong 688- and 871-keV transitions of band 4, the much weaker 827- and 764-keV transitions of band 5, as well as the 333-, 430-, 411-, and 325-keV transitions of band D1. The spectrum in panel (b) of Fig.~\ref{fig2s}, obtained by double-gating on the 688- and 450-keV transitions, shows the strong 747-keV transition and the weaker 871-keV transition in band 4.  This clean spectrum does not show contaminations from other nuclei.  The clean spectrum in panel (c) of Fig.~\ref{fig2s}, obtained by double gating on the 796- and 813-keV transitions, shows the 955-keV transition in band 3 and some of dipole transitions in band D2. 

To extract the asymmetry of the 747-keV transition which connects band 3 to 1, the spectra gated by the 450-keV transition were used. Fig.~\ref{fig3s} shows the asymmetry spectra for the 747-keV transition, in which one can observe the wide peak composed of the 688-keV $E2$ transition in $^{135}$Pr and the 693-keV transition in band N1 of $^{136}$Nd~\cite{136Nd-Lv}, the 721-keV $(20^+\rightarrow18^+)$  $E2$ transition in  $^{134}$Ce  \cite{134Ce-petrache}, and the 772-keV $E2$  in band N1 of $^{136}$Nd~\cite{136Nd-Lv}.
 
To get the polarization of the 813-keV transition linking band 3 to 1, the possible candidate gates are 660 keV (band 1), 796 keV (band 3) and 551 keV (linking band 5 to 4) (see Fig.~\ref{fig1s}). However, the 660-keV transition cannot be used due to the contamination by the strong 663-keV transition in the yrast band of $^{136}$Nd, while the 551-keV transition from band 4 to 3 being very weak, is also useless. Therefore, the only suitable gate is on the 796-keV transition. Fig.~\ref{fig4s} provides the spectra used to extract the polarization asymmetry for the 813-keV transition, obtained by gating on the 796-keV transition. It should be mentioned that the energy of the 796-keV transition is very close to the 794-keV, 17/2$^-$ $\to$ 13/2$^-$ transition in $^{137}$Nd \cite{137nd-npa}, which is also well populated in the present reaction. This is the reason why we also observe the 801-keV, 31/2$^- \to 27/2^-$ and 856-keV, 19/2$^- \to 17/2^+$  transitions in the Fig.~\ref{fig4s}. However, the main contribution to the 856-keV peak is from band HD1 band of $^{136}$Nd~\cite{136Nd-HD}.  Note that there are no 813-keV transitions in $^{136,137}$Nd in coincidence with the 796- or the 794-keV transitions. Thus, the 813-keV peak in the spectrum obtained by gating on the 796-keV transition is clean. 

To obtain the polarization value for the 450-keV transition, the 747-keV transition is the only good gate which ensure sufficient statistics. As can be seen in the double-gated spectrum shown in Fig.~\ref{fig2s}(a), there are no contaminating transitions from other nuclei. However, in the spectrum obtained by gating on only the 747-keV transition shown in Fig.~\ref{fig5s}, the transitions with energies of 407, 437, and 460 keV from $^{137}$Nd are also present. This is due to the contamination from the strong 749-keV, $25/2^-\rightarrow 23/2^-$ transition in $^{137}$Nd \cite{137nd-npa}. The 423-keV transition is from band D4 in $^{135}$Nd which is in coincidence with the 750-keV $21/2^{+}\rightarrow17/2^{+}$ transition~\cite{135nd-lv}. As there are no transitions  of 450 keV in these nuclei in coincidence with the 747-keV transition, the 450-keV peak in the spectrum gated by the 747-keV transition is clean. It should be also mentioned that there is a weak 750-keV, 11$^- \to 10^-$ transition fed by band D3 to band L1 of $^{136}$Nd~\cite{136Nd-Lv}. To avoid contamination from this peak, we used a narrow gate width which eliminate the contamination from the 449-keV, $21^{-} \to 20^{-}$ transition in band D3. As one can see in  Fig.~\ref{fig5s}, the 495-keV, $20^{-} \to 19^{-}$ transition just below the 449-keV transition is not present in the spectrum, which means that there is no contamination from the 449-keV transition of band D3.

The linear polarization and $R_{ac}$ values for the 747-, 813-, and 450-keV transitions are extracted from the spectra presented in Figs.~\ref{fig3s}-\ref{fig5s} by analyzing the polarization matrices using the \textsc{radware}~\cite{rad1, rad2} and \textsc{gaspware} packages \cite{gaspware}. 

The linear polarization $P$ and $R_{ac}$ are a function of mixing ratio $\delta$, and of $\sigma/I$ ($\sigma$ denotes the the variance of the spin projection on the beam direction). Hence, one can uniquely determine $\delta$ by combining the linear polarization $P$ with $R_{ac}$, as shown in Fig.~2 of the main text.  The best fit of the experimental central values of $P$ and $R_{ac}$,  gives the adopted $\delta$ and  $\sigma/I$ values. The calculated $P$ and $R_{ac}$ values for several $\sigma/I$ values were used to fix the upper- and lower- limits of the experimental $P$ and $R_{ac}$ values, which in turn determined the upper- and lower- limits of the adopted mixing ratios.  Note that the adopted error on the mixing ratio is not that induced by the different $\sigma/I$ values: it exactly reflects the error propagation of the uncertainties on $P$ and $R_{ac}$ to the mixing ratio $\delta$.

\begin{table*}[]
 \begin{threeparttable} 
  \centering
  \caption{List of the energies of transitions ($E_\gamma$), excitation energies ($E_i$), intensities (I$_\gamma^b$), anisotropies ($R_{ac}$), multipolarity,  and the spin-parity assignments of the observed states in $^{135}$Pr.}
  \label{tab2}
  \begin{tabular}{cccccc}
 \\
\hline\hline
$E_\gamma^a$ (keV)  &$E_i$(keV)      & I$_\gamma^b$      &$R_{ac}$          &Multipolarity      &J$^{\pi}_i$ $\rightarrow$ J$^{\pi}_f$  \\
\hline
$\rm{Band~1}$ \\  
372.8    &730.6       &100         &1.35(17)                 & E2          &15/2$^{-}$ $\rightarrow$ 11/2$^{-}$     \\
660.2    &1390.8     &78(5)       &1.40(11)                 & E2          &19/2$^{-}$ $\rightarrow$ 15/2$^{-}$     \\ 
844.1    &5164.7     &4.4(4)      & 1.36(18)                  & E2          &35/2$^{-}$ $\rightarrow$ 31/2$^{-}$      \\ 
854.2    &2245.0     &36(2)       & 1.28(13)                & E2          &23/2$^{-}$ $\rightarrow$ 19/2$^{-}$      \\ 
1000.0  &3245.0     &8.4(2)      & 1.33(19)                  & E2          &27/2$^{-}$ $\rightarrow$ 23/2$^{-}$      \\ 
1075.6  &4320.6     &5.1(5)      & 1.35(10)                & E2          &31/2$^{-}$ $\rightarrow$ 27/2$^{-}$      \\ 

$\rm{Band~2}$ \\  
221.3    &951.8        &0.7(2)        &                               &                  &13/2$^{-}$ $\rightarrow$ 15/2$^{-}$      \\
482.1    &1433.9      &2.1(2)        & 1.49(19)                 &E2             &17/2$^{-}$ $\rightarrow$ 13/2$^{-}$      \\
594.0    &951.8        &2.8$^c$     & 0.53(7)                  &M1/E2       &13/2$^{-}$ $\rightarrow$ 11/2$^{-}$      \\
703.2    &1433.9      &1.9(4)        & 0.65(16)                 &M1/E2      &17/2$^{-}$ $\rightarrow$ 15/2$^{-}$      \\ 
725.5    &2159.4      &2.9(3)        & 1.41(23)                 &E2             &21/2$^{-}$ $\rightarrow$ 17/2$^{-}$      \\
768.9    &2159.4      &0.4(2)        &                                &                 &21/2$^{-}$ $\rightarrow$ 19/2$^{-}$      \\ 

$\rm{Band~3}$ \\  
526.2    &1477.9      &0.4(2)     &                              &                   &17/2$^{-}$ $\rightarrow$ 13/2$^{-}$   \\
726.0    &2203.9      &3.7(3)     &1.31(15)                &E2               &21/2$^{-}$ $\rightarrow$ 17/2$^{-}$      \\
747.3    &1477.9      &9.4(5)     &0.37(4)                  &M1/E2         &17/2$^{-}$ $\rightarrow$ 15/2$^{-}$      \\
755.1    &3000.3      &2.1(4)     &0.50(6)                  &M1/E2         &25/2$^{-}$ $\rightarrow$ 23/2$^{-}$      \\
796.4    &3000.3      &3.9(2)     &1.40(10)               &E2               &25/2$^{-}$ $\rightarrow$ 21/2$^{-}$      \\
813.2    &2203.9      &6.4(3)     &0.48(6)                  &M1/E2         &21/2$^{-}$ $\rightarrow$ 19/2$^{-}$      \\
954.9    &3955.2      &1.8(2)     &1.34(12)                &E2              &29/2$^{-}$ $\rightarrow$ 25/2$^{-}$      \\
1112.0   &5067.2      &0.7(1)     &                              &                  &(33/2$^{-})$ $\rightarrow$ 29/2$^{-}$      \\

$\rm{Band~4}$ \\  
412.5    &2616.0      &0.6(3)        &                             &                  &23/2$^{-}$ $\rightarrow$ 21/2$^{-}$      \\
450.2    &1927.7      &4.4(2)        &0.49(4)                &M1/E2        &19/2$^{-}$ $\rightarrow$ 17/2$^{-}$      \\
494.3    &1927.7      &1.1(2)        &                            &                  &19/2$^{-}$ $\rightarrow$ 17/2$^{-}$      \\
688.3    &2616.0      &5.7(2)        &1.42(15)             &E2             &23/2$^{-}$ $\rightarrow$ 19/2$^{-}$      \\
871.1    &3487.1      &2.1(2)        &1.36(17)             &E2              &27/2$^{-}$ $\rightarrow$ 23/2$^{-}$      \\
1197.1  &1927.7      &5.9(5)        &1.39(16)             &E2              &19/2$^{-}$ $\rightarrow$ 15/2$^{-}$      \\ 
1225.5  &2616.0      &3.8(5)        &                            &                  &23/2$^{-}$ $\rightarrow$ 19/2$^{-}$      \\

$\rm{Band~5}$ \\  

518.6    &3519.0      &0.4(1)        &                            &                   &27/2$^{-}$ $\rightarrow$ 25/2$^{-}$      \\
550.9    &2754.6      &0.5(1)        &0.57(10)              &M1/E2         &23/2$^{-}$ $\rightarrow$ 21/2$^{-}$      \\
594.9    &2754.6      &1.0(5)        &                            &                   &23/2$^{-}$ $\rightarrow$ 21/2$^{-}$      \\
764.4    &3519.0      &2.4(3)        &1.34(22)              &E2              &27/2$^{-}$ $\rightarrow$ 23/2$^{-}$      \\
826.9    &2754.6      &1.9(2)       &1.41(14)              &E2              &23/2$^{-}$ $\rightarrow$ 19/2$^{-}$      \\
1364.2  &2754.6      &2.4(4)        &1.27(20)               &E2              &23/2$^{-}$ $\rightarrow$ 19/2$^{-}$      \\

$\rm{Other~states}$ \\  
332.9    &3863.9      &                  &0.65(5)                &M1/E2        &29/2$^{-}$ $\rightarrow$ 27/2$^{-}$      \\
345.1    &3863.9      &1.1(2)        &0.70(10)               &M1/E2        &29/2$^{-}$ $\rightarrow$ 27/2$^{-}$      \\
376.5    &3863.9      &0.7(5)        &                             &                   &29/2$^{-}$ $\rightarrow$ 27/2$^{-}$      \\
530.8    &3531.0      &0.5(2)        &                            &                   &27/2$^{-}$ $\rightarrow$ 25/2$^{-}$      \\
776.4    &3531.0      &3.2(4)       &1.27(18)              &E2              &27/2$^{-}$ $\rightarrow$ 23/2$^{-}$      \\
914.7    &3531.0      & $<$0.5     &                              &                 &27/2$^{-}$ $\rightarrow$ 23/2$^{-}$      \\ 
1286.5  &3531.0      &3.1(3)        &1.32(9)                  &E2              &27/2$^{-}$ $\rightarrow$ 23/2$^{-}$      \\

\hline
\hline
  \end{tabular}
   \begin{tablenotes}    
        \footnotesize              
 \item[a]The error on the transition energies is 0.2 keV for transitions below 1000 keV of the  $^{135}$Pr reaction channel, 0.5 keV for transitions above 1000 keV, and 1 keV for transitions above 1200 keV. \\
\item[b] Relative intensities corrected for efficiency, normalized to the intensity of the 372.8 keV transition. The transition intensities were obtained from a combination of total projection and gated spectra. \\
\item[c] Data was adopted from Ref.~\cite{135Pr1986PRC}.  \\
\end{tablenotes} 
 \end{threeparttable} 
\end{table*}


\bibliography{135Pr}
\end{document}